\documentclass[preprint,showpacs,preprintnumbers,amsmath,amssymb]{revtex4}

\usepackage{graphicx}
\usepackage{dcolumn}
\usepackage{bm}

\begin{document}


\title{The coupling constants for an electroweak model with a $\boldsymbol{SU(4)_{PS} \otimes SU(4)_{EW}}$
unification symmetry\footnote{This work was supported by FAPESP (PD 04/13770-0).}}

\author{A. E. Bernardini}
\affiliation{Department of Cosmic Rays and Chronology, State University of Campinas,\\
PO Box 6165, 13083-970, Campinas, SP, Brazil.}
\email{alexeb@ifi.unicamp.br}

\date{\today}

\begin{abstract}
We introduce the sequence of spontaneous symmetry breaking of a
coupling between Pati-Salam and electroweak symmetries $SU(4)_{PS} \otimes SU(4)_{EW}$ 
in order to establish a mathematically consistent relation among the coupling
constants at grand unification energy scale.
With the values of barion minus lepton quantum numbers of known quarks and leptons, by including right-handed neutrinos,
we can find the mixing angle relations at different energy levels up to the electromagnetic $U(1)_{EM}$ scale.
\end{abstract}

\maketitle

In the context of grand unification theories (GUTs) \cite{Sal68,Moh86,Qui83},
the necessity of going beyond the standard model (SM)
has been established on discussions about gauge coupling unification and
neutrino oscillation \cite{Kay04} as well as on leptogenesis and certain intriguing features
of quark-lepton masses and mixing \cite{Pat03}.
so that a detailed study of the hierarchy problem
has been emphasized and some interesting classes of models have been
proposed \cite{AHD98,PP95,M99,F92,FF99,S80,LP98,L94}.
An extensive class of models has described an electroweak unification scheme
by using a gauge structure with $SU(4)_L$ \cite{PT93} and $SU(3)_L$ \cite{PP92,LP98} symmetries.
Such models introduce some multiplet representations which allow
the existence of quarks and leptons with exotic charges and, in particular, for some of them, 
the right-handed neutrinos appear in the same
representation of these new exotic particles.

By following a similar development, in the place of a $SU(5)$ unification group,
we suggest the coupling between the Pati-Salam (PS) $SU(4)$ symmetry and a {\em left-right} electroweak (EW) $SU(4)$
symmetry described by
\begin{equation}
\begin{array}{lcccc}
SU(4)_{PS} & \otimes &  SU(4)_{EW}&&  \\
   g_{4}     &         &  g_{4} &&   \\
  \Downarrow & \phi_{15-dim}&  &&           \\
 SU(3)_{c}& \otimes &  SU(4)_{EW} &\otimes& U(1)_{B-l}\\
    g_{c}     &         &  g_{4}  &&g_{B-l}  \\
\end{array}
\label{bbb1},
\end{equation}
which could be embedded in a higher symmetry and, therefore, make us able to find a relation between the $g_4$
and the $g_{B-l}$ coupling constants.
As a first analysis, we have supposed that a Higgs boson designated by a $15-dim$ representation of $SU(4)_{PS}$
leads to the spontaneous symmetry breaking (SSB) of $SU(4)_{PS}$ into $SU(3)_{c} \otimes U(1)_{B-l}$
where the color number is represented by $c$ and the barion minus lepton number by $B-l$.
By including the first generation of the model in the following matrix representation,
\begin{equation}
\Psi = \left[
\begin{array}{llll}
u_{1L}  & u_{2L}  & u_{3L}  & \nu^e_{L}  \\
d_{1L}  & d_{2L}  & d_{3L}  & e_{L}      \\
u_{1R}  & u_{2R}  & u_{3R}  & \nu^e_{R}  \\
d_{1R}  & d_{2R}  & d_{3R}  & e_{R}
\end{array}
\right].
\label{bbb3}
\end{equation}
and assuming a universal coupling constant $g_G$ in a way that we can establish
\begin{equation}
G \supset SU(4)_{PS}  \otimes   SU(4)_{EW}
\label{bbb3b},
\end{equation} 
we can express the the free Lagrangian as
\begin{equation}
\mathcal{L} = Tr \left[\overline{\Psi} \gamma^{\mu} \mbox{\boldmath$D_{\mu}$} \Psi \right] + \mbox{coupling and interactions}
\label{bbb3c},
\end{equation}
with
\begin{equation}
\mbox{\boldmath$D_{\mu}$} \Psi =  \partial _{\mu} \Psi - i g_G \left( h^a_{\mu (EW)} H^a \Psi + h^b_{\mu (PS)} \Psi H^b \right)
\label{bbb31}
\end{equation}
where $H^{a}$ and $H^{b}$ are the $SU(4)$ generators for, respectively, $EW$ and $PS$ interactions, which
are not exactly in the same irreducible representation.
The above assumptions allow us to establish a consistent relation among coupling constants at grand unification energy scale that could be extended to phenomenological constraints at electroweak scale.
To analyze all the possibilities of SSB
relative to the $SU(4)_{EW}$ symmetry, we analyze three cases with a minimum number of Higgs bosons\footnote{Obviously they are not sufficient for providing mass to all fermions and gauge bosons.}
necessary to obtain the final electromagnetic symmetry $U(1)_{EM}$\\

\paragraph*{Case 1\\}

\begin{equation}
\begin{array}{ccc}
SU(4)_{EW}   & \otimes & U(1)_{B-l} \\
   g_{4}     &         &  g_{B-l}    \\
  \Downarrow &\chi_{4}&             \\
SU(3)_{L(R)} & \otimes & U(1)_{Y_{SU(3)}}  \\
   g_{3}     &         &  g_{Y_{SU(3)}}    \\
  \Downarrow &\chi_{3}&                    \\
SU(2)_{L}    & \otimes & U(1)_Y      \\
   g         &         &  g^{\prime} \\
  \Downarrow &\rho,   \eta&                  \\
U(1)_{EM}    &         &  \\
   e         &         &  \\
\end{array}
\label{bbb2a}.
\end{equation}
where $\chi_n$ are $n-dim$ multiplets and $\rho$ and $\eta$ are Higgs doublets which provide
the successive SSB.
We can obtain the neutral Hermitian gauge boson mass matrix $\mathbf{M}^2$ written in the ${\mathbf W}$ basis of $SU(4)_{EW}$ diagonal generators
\begin{equation}
\Delta \mathcal{L}_{mass}=\frac{1}{2} {\mathbf W}^{\dagger}\cdot {\mathbf M}^2 \cdot{\mathbf W}
\label{bb1},
\end{equation}
with
\begin{equation}
{\mathbf W}^{\dagger} = \left(
\begin{array}{cccc}
h_\mu^3 & h_\mu^8 &  h_\mu^{15} & d_\mu
\end{array}\right),
\label{bb2}
\end{equation}
and
\begin{equation}
\begin{array}{lcr}
H_3     &= & \frac{1}{2}         ~diag \left[1,   -1,  ~~~~0,   ~~~~0\right],\\
&&\\
H_8     &= & \frac{1}{2\sqrt{3}} ~diag \left[1,~~~~1,     -2,   ~~~~0\right],\\
&&\\
H_{15}  &= & \frac{1}{2\sqrt{6}} ~diag \left[1,~~~~1,  ~~~~1,      -3\right].
\end{array}
\label{bb3}
\end{equation}
The gauge bosons which carry the quantum numbers of the $U(1)_{EM}$, $U(1)_Y$ and $U(1)_{Y_{SU(3)}}$ symmetries are calculated
and represented respectively by
\begin{eqnarray}
A_\mu &=& \frac{1}{\sqrt{1+2t^2}}\left(t~h_\mu^3 \mp \frac{t}{\sqrt{3}}~ h_\mu^8 \pm t \sqrt{\frac{2}{3}} ~h_\mu^{15} + d_\mu \right),
\end{eqnarray}
\begin{eqnarray}
B_\mu~~~~&=& \frac{1}{\sqrt{1+t^2}}\left( \mp \frac{t}{\sqrt{3}}~h_\mu^8 \pm t \sqrt{\frac{2}{3}} ~h_\mu^{15} + d_\mu \right),
\end{eqnarray}
\begin{eqnarray}
B_\mu^{SU(3)} &=& \sqrt{\frac{3}{3+2t^2}}\left( \pm  t \sqrt{\frac{2}{3}} ~h_\mu^{15} + d_\mu \right),
\end{eqnarray}
where $d_\mu$ is the gauge boson related to the $U(1)_{B-l}$ symmetry and
the parameter $t$ is defined as $t = \frac{g_{B-l}}{g_{4}}$.
In this way, the relation among $t$ and the mixing angles can be established by
\begin{equation}
\begin{array}{llrccccllrcc}
\sin{\theta_w}&= &\frac{t}{\sqrt{1+2t^2}}       &&& & &\cos{\theta_w} &= &\sqrt{\frac{1+t^2}{1+2t^2}}, &&\\
\sin{\theta_3}&= &-\frac{t}{\sqrt{3+3t^2}}   & &&& &\cos{\theta_3} &= &\sqrt{\frac{3+2t^2}{3+3t^2}}, &&\\
\sin{\theta_4}&= &\sqrt{\frac{2t^2}{3+2t^2}} & &&& &\cos{\theta_4} &= &\sqrt{\frac{3}{3+2t^2}},&&
\end{array}
\label{bb9}
\end{equation}\\

\paragraph*{Case 2\\}
\begin{equation}
\begin{array}{ccccc}
SU(4)_{EW}   & \otimes & U(1)_{B-l} \\
   g_{4}     &         &  g_{B-l}    \\
  \Downarrow &\Phi_{15}&            \\
SU(2)_{L}    & \otimes & SU(2)_{R} & \otimes &  U(1)_{B-l}  (\otimes  U(1)_{EW})    \\
   g_{L}     &         &     g_{L} &         &    g_{B-l}~~~~~~~~~~~~~~~~    \\
  \Downarrow & \Phi_{8}&                   \\
SU(2)_{L}    & \otimes & U(1)_Y (\otimes  U(1)_{EW})     \\
   g         &         &  g^{\prime}~~~~~~~~~~~~~~~~ \\
  \Downarrow &\rho     &             \\
U(1)_{EM}    &         &  \\
   e         &         &  \\
\end{array}
\label{bbb2b}.
\end{equation}
where $\Phi_n$ are $n-dim$ multiplets and $\rho$ is a Higgs doublet which provide
another sequence of successive SSB.
We can obtain, again, the neutral Hermitian gauge boson mass matrix $\mathbf{M}^2$, but now, written in another ${\mathbf W}$ basis of diagonal generators
of $SU(4)_{EW}$,
\begin{equation}
\begin{array}{lcr}
H_3     &= & \frac{1}{2}         ~diag \left[1,   -1,  ~~~~0,   ~~~~0\right],\\
&&\\
H_8     &= & \frac{1}{2}         ~diag \left[0,~~~~0,  ~~~~1,      -1\right],\\
&&\\
H_{15}  &= & \frac{1}{2\sqrt{2}} ~diag \left[1,~~~~1,     -1,      -1\right].
\end{array}
\label{bb10}
\end{equation}
The gauge bosons related to the $U(1)_{EM}$ and $U(1)_Y$ symmetries are respectively
\begin{eqnarray}
A_\mu &=& \frac{1}{\sqrt{1+2t^2}}\left(t~h_\mu^3 + t~h_\mu^8 + d_\mu \right),
\end{eqnarray}
\begin{eqnarray}
B_\mu &=& \frac{1}{\sqrt{1+t^2}}\left(t~h_\mu^8  + d_\mu \right).
\end{eqnarray}
In this case, the relation among the parameter $t$ and the mixing angles can be given by
\begin{equation}
\begin{array}{lllcclllcc}
\sin{\theta_w}&= &\frac{t}{\sqrt{1+2t^2}}  & & &\cos{\theta_w} &= &\sqrt{\frac{1+t^2}{1+2t^2}}, &&\\
\sin{\theta_R}&= &\frac{t}{\sqrt{1+t^2}}   & & &\cos{\theta_R} &= &\frac{1}{\sqrt{1+ t^2}}, &&\\
\end{array}
\label{bb12}
\end{equation}\\

\paragraph*{Case 3\\}
\begin{equation}
\begin{array}{ccccc}
SU(4)_{EW}   & \otimes & U(1)_{B-l} \\
   g_{4}     &         &  g_{B-l}    \\
  \Downarrow &\Phi_{8}&             \\
SU(2)_{L}    & \otimes & U(1)_Y (\otimes  U(1)_{EW})     \\
   g         &         &  g^{\prime}~~~~~~~~~~~~~~~~ \\
  \Downarrow &\rho     &             \\
U(1)_{EM}    &         &  \\
   e         &         &  \\
\end{array}
\label{bbb2c}.
\end{equation}
Here we immediately notice a simplification of the second case where the same mixing angle relations can be reproduced.
At this point, it is pertinent to observe that the $U(1)$ symmetry that
appears after the SSB of $SU(4)_{PS} \otimes SU(4)_{EW}$ carries the $B - l$ quantum numbers
which provide the correct values for the electric
charge $Q$ quantum numbers to all fermions and bosons.
The three cases suggest the possibility of obtaining mixing angle values by introducing a numerical value for $t$.
Proceeding with the unification analysis, independently of the interactions or symmetry breaking mechanisms, the
set of generators $\tau^a$ of a unification symmetry (group)
must agree with the following {\em trace} relation
\begin{equation}
Tr[\tau^a \tau^b] = C \delta^{ab}
\label{merer},
\end{equation}
where $C$ has the same value for each subgroup even depending on the representation over which the traces are taken.
To exemplify this point for the above cases, let us observe that the kinetic term (\ref{Higg2}) related to the Pati-Salam symmetry
can be written as
\begin{equation}
\Delta\mathcal{L}_{kinetic} = \overline{\psi^a_{PS}}\gamma^{\mu} D_{\mu}\psi^a_{PS}
\label{Higg2},
\end{equation}
with the left (right) chiral spinor fields $\psi^i_{PS}$ represented by
\begin{equation}
\psi^{1(3)}_{PS} = \left(\begin{array}{l} u_{r}\\ u_{g}\\ u_{b} \\ \nu_e \end{array}\right)_{L(R)} ~~~~ \mbox{and} ~~~
\psi^{2(4)}_{PS} = \left(\begin{array}{l} d_{r}  \\ d_{g}    \\ d_{b}    \\    e  \end{array}\right)_{L(R)},
\label{pspsps}
\end{equation}
where $r$, $g$ and $b$ are the color index.
The Pati-Salam covariant derivative thus given by
\begin{equation}
D_\mu =  \partial _\mu -  i g_{4} h_{\mu (PS)}^a H^a
\label{m6},
\end{equation}
which after a SSB provided by a $15-dim$ Higgs boson becomes
\begin{equation}
D_\mu =  \partial _\mu - i g_cG^b_\mu \frac{\lambda^b}{2} - i g_{B-l}d_{\mu} \frac{B - l}{2} - i g_{4}\sum_{i=9}^{14}(h^a_{\mu}H^a)
\label{m234},
\end{equation}
where $G_\mu^{a}$ are related to the massless gluons of $SU(3)_{c}$, $d_{\mu}$ is
related to the massless gauge boson that keeps the hypercharge invariance $B-l$ of $U(1)_{B-l}$.
and last term could be rewritten in terms of the mass eigenstates which appears after SSB.

Turning back to the Eq.~(\ref{merer}) and taking the $4-dim$ multiplet $\psi^i_{PS}$ of Eq.~(\ref{pspsps}), we obtain
\begin{equation}
\begin{array}{lclcl}
Tr[g^2_{4} H_a^2] &  & = &  & \frac{1}{2} g^2_{4},
\end{array}
\end{equation}
when we consider the normalized generators $H_a$ of $SU(4)_{PS}$.
By the same way,
\begin{equation}
\begin{array}{lclcl}
Tr[g^2_{(c)}\lambda_a^2] & = & \frac{1}{2} \left[\frac{1}{2} + 0 + \frac{1}{2} + 0\right] &=& \frac{1}{2} g^2_c,\\
\end{array}
\end{equation}
when it is applied  to the normalized generators $\lambda_a$ of $SU(3)_{c}$ and summed over a color triplet (quarks) and a color singlet (lepton).
Finally,
\begin{equation}
\begin{array}{lclcl}
Tr[g^2_{B-l} \left(\frac{B - l}{2}\right)^2] & = & \frac{1}{4}\left[\frac{1}{9} + \frac{1}{9} +\frac{1}{9} +1\right] &=& \frac{1}{3} g^2_{B-l},
\end{array}
\end{equation}
which is constructed in terms of the hypercharge of $U(1)_{B-l}$ summed over three quarks and one lepton.
Once we have previously assumed
\begin{equation}
G_{Unif} = g_4 = g_3 = g_L = g_R
\label{bbb4},
\end{equation}
we are now in position to determine the relation between $g_4$ and $g$ or $g^{\prime}$.
From Eq.~(\ref{merer}) we may write
\begin{equation}
g_{4} = g_{c} = \sqrt{\frac{2}{3}} g_{B-l}.
\label{acopl}
\end{equation}
and, by summarizing the above procedure, when we make $t = \sqrt{\frac{2}{3}}$ in agreement with
the last equality, the mixing angles and the coupling constant relations can be written as
\begin{equation}
\begin{array}{rrrrrrrrrr}
~~~~s{\theta_w}&=& \sqrt{\frac{3}{8}}     & ~~~~c{\theta_w} &=& \sqrt{\frac{5}{8}} &  ~~~~g^{\prime}   &=&  \sqrt{\frac{3}{5}}g   &\\
~~~~s{\theta_3}&=&  \frac{1}{\sqrt{5}}    & ~~~~c{\theta_3} &=& \frac{2}{\sqrt{5}} &  ~~~~g_{Y_{SU(3)}}&=&  \frac{1}{2} g_3       &\\
~~~~s{\theta_4}&=& -\frac{\sqrt{2}}{2}    & ~~~~c{\theta_4} &=& \frac{\sqrt{2}}{2} &  ~~~~g_{B-l}       &=& \sqrt{\frac{3}{2}} g_4 &\\
~~~~s{\theta_R}&=& \sqrt{\frac{3}{5}}     & ~~~~c{\theta_R} &=& \sqrt{\frac{2}{5}} &  ~~~~g_{B-l}       &=&  \sqrt{\frac{3}{2}}g_R &
\end{array}
\label{bb20}
\end{equation}
where $s{\theta} \equiv \sin{\theta}$ and $c{\theta} \equiv \cos{\theta}$.
Obviously, the electroweak mixing angle $\theta_w$ has the same value in all the three cases
and corresponds to one of few experimental concordant results generated by the $SU(5)$ model \cite{GG74}.
Furthermore, the model concerns about the existence of a right-handed neutrino as a fundamental lepton so that
the discussion about anomaly cancellation is unnecessary since, for each fermion generation,
a {\em left-right} symmetry is assumed.
All the elements are constructed without introducing any exotic character and, with some phenomenological 
analysis, these results could be extended to the study of gauge boson mass generation and neutral and charged vector currents.

\end{document}